# Interfacial Charge Transfer and Ultrafast Photonics Application of 2D Graphene/InSe Heterostructure

Jialin Li[1], Lizhen Wang[2], Yuzhong Chen[2], Yujie Li[2], Haiming Zhu[2], Linjun Li[1,3,*] and Limin Tong[1,3]

1. State Key Laboratory of Modern Optical Instrumentation, College of Optical Science and Engineering, Zhejiang University, Hangzhou 310027, China. 11830036@zju.edu.cn; 11530046@zju.edu.cn; lilinjun@zju.edu.cn; phytong@zju.edu.cn.
2. Center for Chemistry of High-Performance & Novel Materials, Department of Chemistry, Zhejiang University, Hangzhou 310027, China. 21637009@zju.edu.cn; 3140105741@zju.edu.cn; hmzhu@zju.edu.cn.
3. Intelligent Optics & Photonics Research Center, Jiaxing Research Institute Zhejiang University, Jiaxing 314000, China.
* Correspondence: lilinjun@zju.edu.cn.

**Abstract:** Interface interactions in 2D vertically stacked heterostructures play an important role in optoelectronic applications, photodetectors based on graphene/InSe heterostructures had shown promising performance nowadays. However, nonlinear optical properties studies based on the graphene/InSe heterostructure was insufficient. Here, we fabricated graphene/InSe heterostructure by mechanical exfoliation, and investigated the optically induced charge transfer between graphene/InSe heterostructures by taking photoluminescence and pump-probe measurements. The large built-in electric field at the interface is confirmed by Kelvin probe force microscopy. Furthermore, due to the efficient interfacial carrier transfer driven by built-in electric potential (~ 286 meV) and broadband nonlinear absorption, the application of graphene/InSe heterostructure in mode-locked laser is realized. Our work not only provides a deeper understanding for the dipole orientation related interface interactions on the photoexcited charge transfer of graphene/InSe heterostructure, but also enrich the saturable absorber family for ultrafast photonics application.

## 1. Introduction

Two-dimensional (2D) heterostructures which combined by van der Waals (vdWs) force, have attracted great attentions for next-generation optoelectronic devices, because of excellent physical properties, such as strong light–matter interactions and ultrafast interfacial charge transfer[1-5]. Graphene, as a well-known 2D material, has been widely used in optoelectronic area due to its ultrafast electron relaxation time[6,7]. However, the shortcoming of graphene for optoelectronic application is the relatively low absorption[8,9]. Recently, an III–VI group layered semiconductor InSe, also shows great interest in optoelectronics area, with high carrier mobility[10], tunable bandgap[10,11] and high nonlinear absorption coefficient[12,13]. For example, InSe has been demonstrated as broadband photodetectors[14,15], saturable absorber (SA) in ultrafast fiber lasers[16-19] and solid-state bulk lasers[20]. Although photodetectors based on graphene/InSe vdWs heterostructure operating at the visible to near infrared (NIR) wavelength range have been reported[21-24], they have been demonstrated the high photodetection performance for phototransistors application due to the suitable band alignment and interfacial charge transfer, all the measurements presented in previous work were obtained under quasi-static conditions. To probe the role of dipole orientation related interface interactions on the dynamic photoexcited charge transfer process in graphene/InSe heterostructure, ultrafast pump-probe optical spectroscopy studies are required. In addition, 2D graphene/InSe based heterostructure for nonlinear photonic application haven't been studied, it is expectable that the heterostructure combines both advantages of ultrafast relaxation and a large effective nonlinear absorption coefficient for higher performance.

In this work, we prepared the graphene/InSe (G/InSe) heterostructure (HS) by mechanical exfoliation (ME) and investigated the intrinsic interlayer charge transfer process of G/InSe HS by steady-state photoluminescence (PL), Kelvin probe force microscopy (KPFM) and transient absorption (TA) pump-probe measurements. We further integrated the G/InSe HS as a SA into an Erbium-doped fiber laser for near-IR mode-locked pulse application. Stable traditional soliton pulses are obtained with central wavelength of 1566.65 nm and 192 fs pulse duration. These results indicate that G/InSe HS is very attractive for ultrafast nonlinear photonic applications.

## 2. Preparation and Characterization

Our Bridgman method grown InSe crystals are obtained commercially (from SixCarbon Technology). The XRD pattern of bulk InSe is shown in Figure 1a, demonstrating that the crystal structure is β phase, which belongs to the space group P63/mmc, and the lattice parameters are a = 4.05 Å, b = 4.05 Å, c = 16.93 Å respectively. Raman scattering measurement on InSe bulk single crystal is performed by a commercial Raman spectrometer (Witech alpha300) with a $\lambda$=532 nm laser for excitation. As shown in Figure 1b, there exists three Raman-vibration modes, including $A_{1g}^1$ (115 cm$^{-1}$), $E_{2g}^1$ (176 cm$^{-1}$), and $A_{1g}^2$ (226 cm$^{-1}$), similar to the result of ref[25], which proves that the sample is β phase with high single crystalline quality. Figure 1c shows a typical surface topographic image taken by scanning electron microscopy (SEM), the corresponding elemental ratio of InSe is obtained by the energy-dispersive X-ray (EDX) spectroscopy, which reveals an In:Se ratio of 1.16 as shown in Figure 1d. The elemental mapping of In and Se are shown in Figure 1e and 1frespectively, indicating of the uniform elemental spatial distribution.

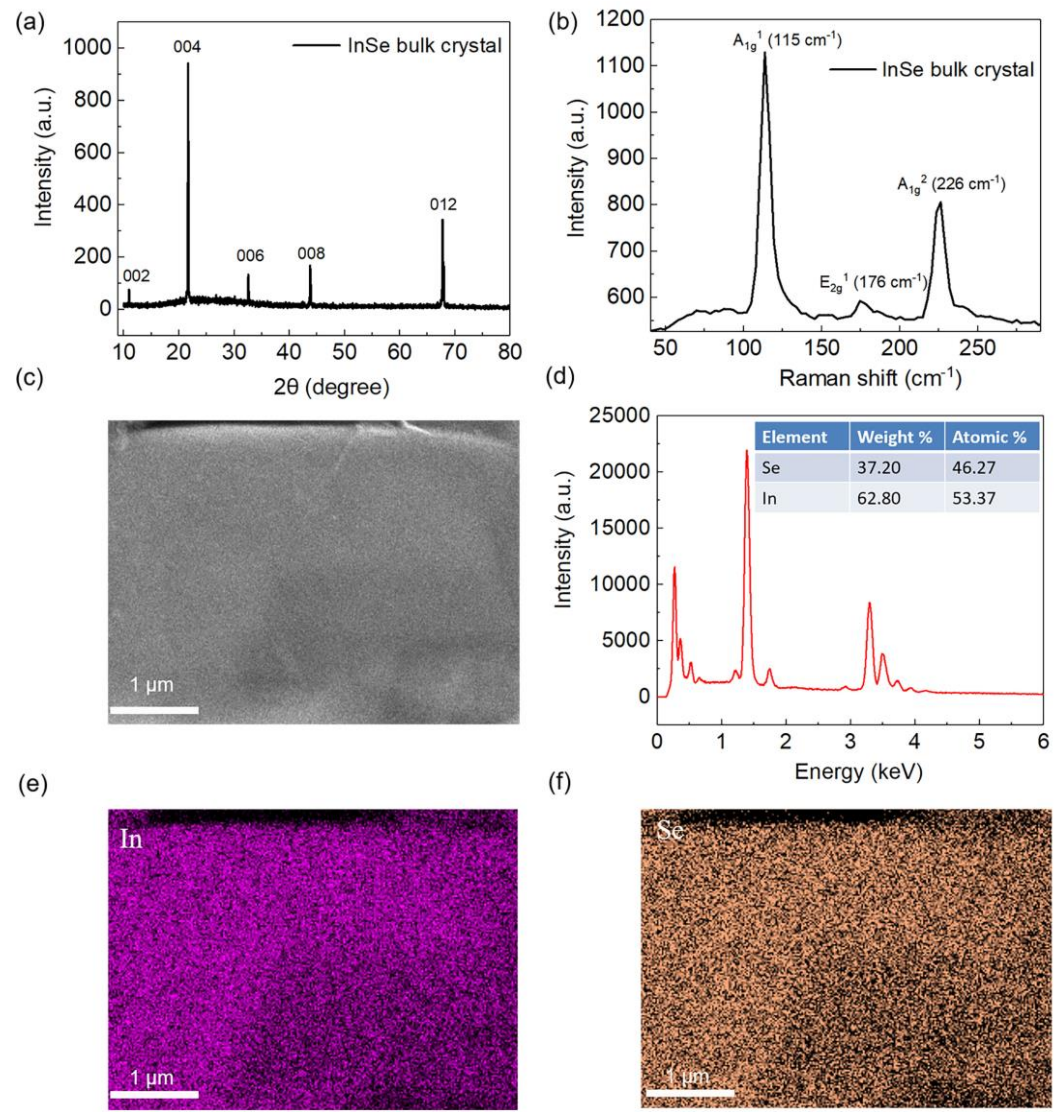

**Figure 1.** (a) XRD spectrum of InSe bulk crystal. (b) Raman spectrum of InSe bulk crystal. (c) SEM image of InSe bulk crystal. (d) EDX of InSe bulk crystal. (e) and (f) Elemental mapping image of InSe bulk crystal.

For fabrication of graphene/InSe heterostructure, it can be divided into three steps. Firstly, the thin flake of graphene was prepared by ME method using Nitto tape [26], and transferred on quartz substrate. Secondly, the thin flake of InSe was firstly exfoliated on PDMS [27] and then transferred on top of the part of the graphene flake assisted by a transfer station under a Nikon microscopy. Thirdly, the sample was then thermal annealed in Ar atmosphere under 200 ℃ for 2 hours to remove chemical residues and improve the interfacial contact. Figure 2a depicts the optical image of the G/InSe and its Raman spectrum(derived by a commercial Raman spectrometer Witech 300R) is shown in Figure 2b. The PL measurements were also performed on G/InSe samplederived by the same Raman spectrometer, using a 532 nm laser for excitation with the power of 700 μW. As shown in Figure 2c, PL is quenched in the heterostructure region, indicating the separation of photoexcited electron–hole pairs occurs at the G/InSe interface.

To determine the work function difference of graphene and InSe, Kelvin probe force microscopy (KPFM) was used. According to the equation $V = (W_{sample} − W_{tip})/e$, the difference in the work function (W) stands for the variation of the surface potential (V) of the sample. Figure 2d shows the KPFM curve of graphene and G/InSe. The result indicates that graphene is heavily p-type doped, while the InSe is n-type doped, leading to the formation of a p–n junction with a large built-in electric potential (the potential difference is~ 0.286 eV). Figure 2e shows the surface Work function mapping image of graphene and G/InSe. Thus, electron transfer is from graphene to InSe, enhancing the p-type carrier concentration in graphene and n-type concentration in InSe. The band alignment of the heterostructure is illustrated in Figure 2e according to the work function. The unique band alignment of G/InSe HS permits the possible transition from graphene to InSe under a low photon-energy excitation, which can be used as a SA in C/L-band pulsed lasers. In addition, the large built-in electric field existing in the G/InSe heterostructure, can inhibit electron–hole recombination and reduce the rate of recombination, leading to a fast electron relaxation time, which is benefit for achieving ultrafast laser pulse.

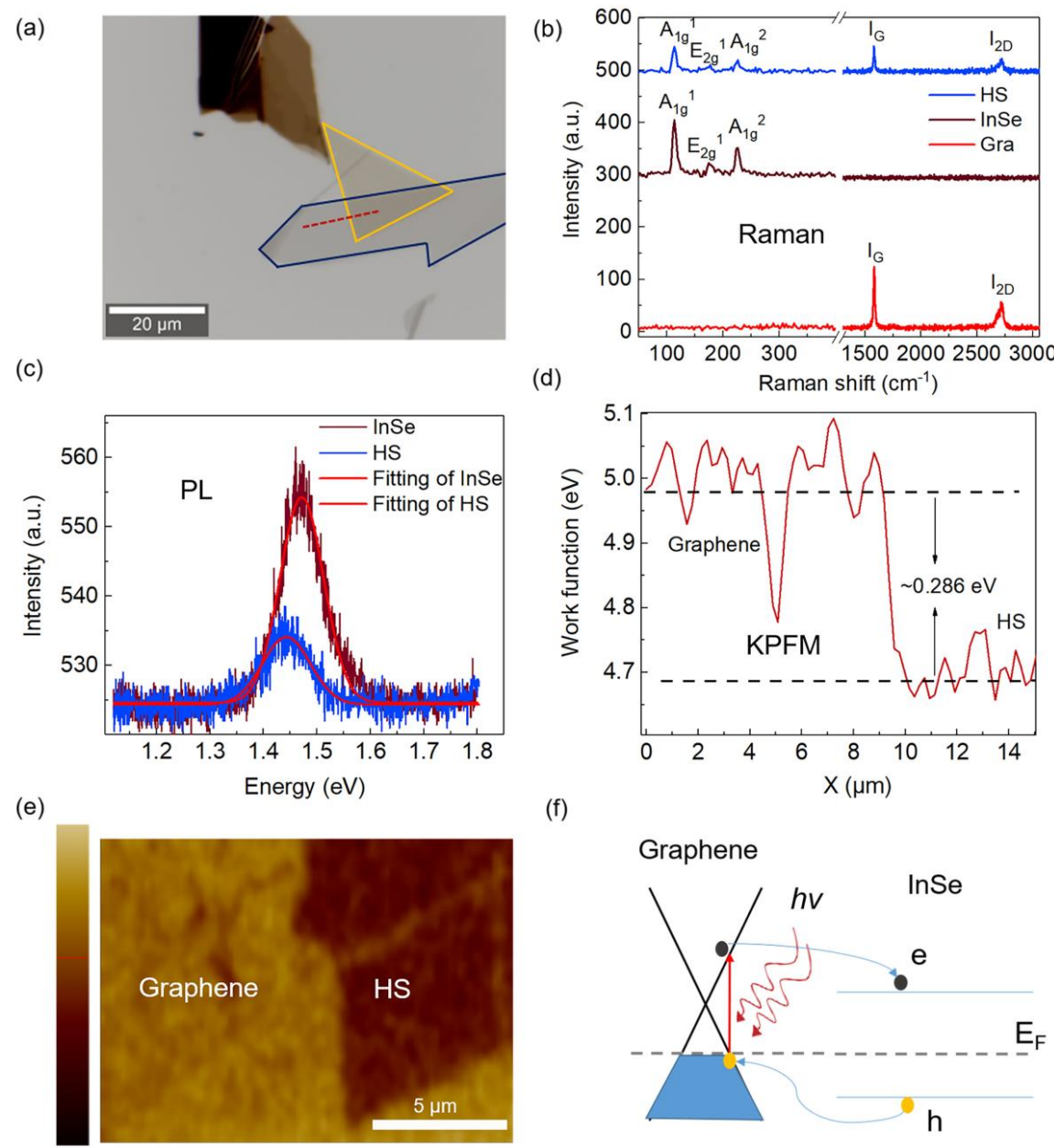

**Figure 2.** (a) Optical image (The blue line is bottom graphene and the yellow line is top InSe), (b) Raman spectrum, (c) PL spectrum of graphene, InSe and G/InSe HS. (d) Work function difference of graphene and HS (corresponding to the red line region in Figure 2a). (e) KPFM surface Work function image of graphene and HS. (f) The band alignment of the heterostructure.

### 3. Carrier Dynamics

To probe carrier transport process across the G/InSe interface and reveal the ultrafast nonlinear optical properties of the G/InSe in visible-near infrared (IR) spectral region, a micro-area pump–probe technique is employed. The TA measurement setup is shown in Figure 3a. . Specifically, the output fundamental beam ($\lambda$=1030 nm, ~170 fs pulse duration) is split into two paths from a Yb: KGW laser (Light Conversion Ltd Pharos), one is sent into a noncollinear optical parametric amplifier (OPA) to produce pump pulse at near ultraviolet, visible and near-IR wavelengths, and the other is focused onto a YAG crystal after a delay line to generate probe pulse of white light continuum ($\lambda$=500 ~ 950 nm) or near-IR ($\lambda$=1425 ~ 1600 nm) light. The pump and probe beams are recombined and focused on the sample through a reflective 50× objective. The femto laser spot size is about 2 μm.

The transient absorption (TA) signal of InSe mainly comes from free carriers, such carriers are the free holes of the valence band (VB1 and VB2) level. Photo-bleaching signature arising from Pauli blocking indicates the saturable absorption behavior of the prepared samples. Under excitation by a 2.06 eV, ~130 μJ cm$^{-2}$ pump pulse, the TA dynamics of graphene, InSe and G/InSe HS is shown in Figure 3b and Figure 3c, which A exciton (844 nm) and B exciton (502 nm) of InSe are selected as the probe wavelength respectively. Fitting the dynamics of them, the G/InSe HS shows a biexponential decay with an intra-band relaxation time ($\tau_1$=48 fs), inter-band relaxation time ($\tau_2$=140 fs), which is closer to graphene ($\tau_1$=96 fs and $\tau_2$=583 fs), much faster than InSe ($\tau_1$=1.77 ps and $\tau_2$=216.11 ps) at probe wavelength of 844 nm as Figure 3a. A exciton of InSe is related to the transition across the principal band gap, which has a dominantly electric dipole-like character, coupled to out-of plane polarized photons[28], so it has less absorption in our configuration which the laser polarization is in-plane (as shown in insert of Figure 3a). In contrast, the B exciton is related to the transition which is strongly coupled to in-plane polarized light. At probe wavelength of 502 nm in Figure 3c, the TA signal is attributed to the two-photon absorption, the relaxation time of G/InSe HS is determined to be $\tau_1$=12.6 ps and $\tau_2$=12.6 ps, which is also faster than InSe individual ($\tau_1$=3.8 ps and $\tau_2$=191.3 ps). Benefit from the fast charge transfer and relaxation channel in graphene, the carrier recombination in InSe is suppressed, and thus results a short relaxation time.. Furthermore, TA dynamics of InSe and G/InSe HS under the pump photon energy of 3.30 eV, ~2 uJ/cm$^2$ is shown in Figure 3d. Compared to InSe, the relaxation time of G/InSe HS is faster than InSe itself, indicating the fast charge transfer at the interface of G/InSe HS. In addition, the TA signal is stronger compared with the signal under 2.06 eV, ~130 μJ cm$^{-2}$ at the probe wavelengths of 502 nm (B exciton), therefore it can be confirmed as the two-photon absorption at the pump photon energy of 2.06 eV. Overall, the ultrafast carrier dynamics revealed the dipole orientation related interface interactions in G/InSe HS, the results not only provide a deeper understanding on the dynamic photoexcited charge transfer process, but also suggest a high optical modulation speed for nonlinear photonics application.

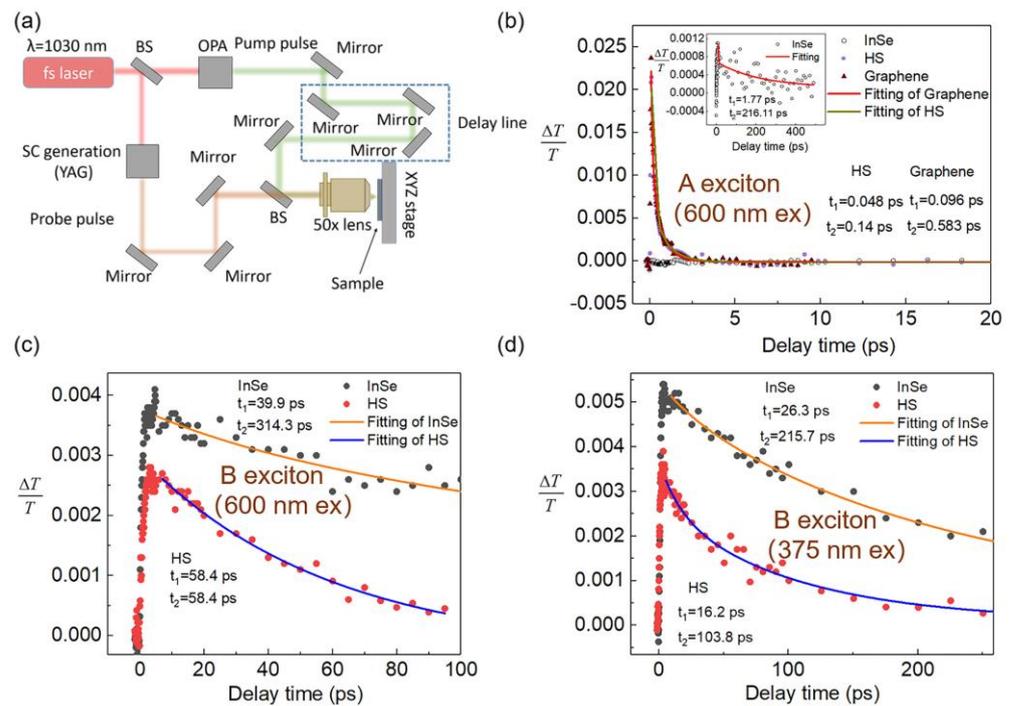

**Figure 3.** (a) Schematic diagram of the micro-area TA setup. (b) TA dynamics of graphene, InSe and G/InSe HS with pump and probe wavelengths of 600 nm and 844 nm (A exciton), respectively. (c) TA dynamics of InSe and G/InSe HS with pump and probe wavelengths of 600 nm and 502 nm (B exciton), respectively. (d) TA dynamics of InSe and G/InSe HS with pump and probe wavelengths of 375 nm and 502 nm (B exciton), respectively.

## 4. Nonlinear Saturable Absorption and Mode-locked Fiber Laser Applications

To measure the saturable absorption properties of G/InSe HS, the micro I-scan with a balanced twin-detector measurement method is used, as illustrated in Figure 4a. The pulse laser source is worked at 1550 nm with 300 fs pulse duration and 100 kHz repetition rates. The sample was located at the focal point of the laser through a 50x objective lens, and an adjustable attenuation filter was used to modulate the laser power. As shown in Figure 4b, the obtained transmittance data are fitted by using the two-level saturable absorption model with the equation $\alpha = \alpha_{ns} + \alpha_s/(1 + I/I_s)$[29]. The saturable intensity and modulation depth obtained are 1.33 GW/cm$^2$ and 12 % respectively. It should be noted that the large modulation depth and low saturation intensity are attributed to the lower recombination rate caused by the large built-in electric potential[30].

To evaluate its ultrafast nonlinear optical response, the G/InSe HS SA was inserted into a ring fiber cavity. Figure 4c shows the schematic of the all-fiber G/InSe HS ring laser. To pump a 0.4-m-long erbium-doped gain fiber (EDF, Liekki Er110-4/125), a commercial continuous wave (CW) 980 nm laser diode (LD) with maximum power of 800 mW was used as pump source. To couple the pump laser into the cavity and prevent back reflection for ensuring unidirectional laser operation, a 980/1550 nm wavelength division multiplexer/isolator hybrid (WDM+ISO) was used. A polarization controller (PC) is used to tune the laser polarization state in the cavity, and the cavity also comprises a 5.2 m-long single mode fiber (SMF-28e). The dispersion parameters of EDF and SMF are 12 ps$^2$/km and -23 ps$^2$/km respectively, therefore the calculated net cavity dispersion is about -0.106 ps$^2$. The laser was output through a 10% optical coupler (OC), and its characteristics is real-time monitored by a 1 GHz photodetector (Thorlabs DET01CFC), and then output to 3-GHz digital oscilloscope (LeCroy WavePro7300). Also, a commercial optical spectrum analyzer (Yokogawa AQ6370D) is used to record the obtained laser optical spectrum. The obtained laser pulse width is measured by using a commercial autocorrelator (APE

Pulsecheck-USB-50), and the radio frequency spectrum is measured by a RF spectrum analyzer (Rigol DSA1030).

By carefully controling the laser polarization state through PCs, self-starting mode-locking is achieved when the pump power is beyond 40 mW, due to the saturable absorption of G/InSe HS SA. The SA damage threshold is about 650 mW. The maximum output laser power is about 2.53 mW, corresponding to the single pulse energy of 0.06 nJ. The output pulse characteristics under the pump power of 110 mW are shown in Figure 4d and Figure 4e. Figure 4c shows the typical output optical spectrum with several pairs of Kelly sidebands, which indicates the signature of conventional soliton operation. The spectrum shows 3-dB bandwidth of 7.43 nm centered at 1566.65 nm. Figure 4e presents the typical output pulse trains. The time interval of mode-locked pulse is about 24.41 ns, well-matched with total cavity length. The RF spectrum around the fundamental repetition rate of 40.96 MHz with the signal-to-noise ratio (SNR) of ~27 dB is shown in inset of Figure 4e, which indicates good stability of the G/InSe HS mode-locked laser. The G/InSe HS SA is stable over two weeks at ambient environment and the mode-locking operation can be stable for over 10 hours. The output mode-locked pulses are amplified by a home-made Erbium-doped fiber amplifier (EDFA) and compressed by a piece of dispersion compensating fiber. As it is illustrated in Figure 4f, the obtained mode-locking pulse duration is about 270 fs.. By fitting the autocorrelation (AC) trace with Gaussian function, the actual pulse duration is estimated to be about 192 fs.

Furthermore, we compared the mode-locked laser results with naked graphene (before stacked with InSe) SA, it shows that the mode-locking pulse width is about 292 fs (Figure S1), slightly slower than G/InSe HS, due to the slower electron relaxation time proved by TA measurement at the probe wavelength of 1566 nm (Figure S2). The internal mechanism is the reduced recombination rate, and fast interlayer electron transfer due to the large built-in electric field in G/InSe HS. Besides, InSe shows a large effective nonlinear absorption coefficient ($\beta_{eff}$ ~ -2.8 × $10^2$ cm/GW)[13], while the heterostructure possesses larger nonlinear absorption coefficient[31]. Furthermore, the ultrafast electrons transfer from graphene to other 2D semiconductors occur from visible to mid-infrared region[32]. Benefiting from the ultrafast relaxation time and strong broadband nonlinear absorption, we believe that the heterostructure could be realized for ultrafast broadband nonlinear optical applications, not only limited to C/L band.

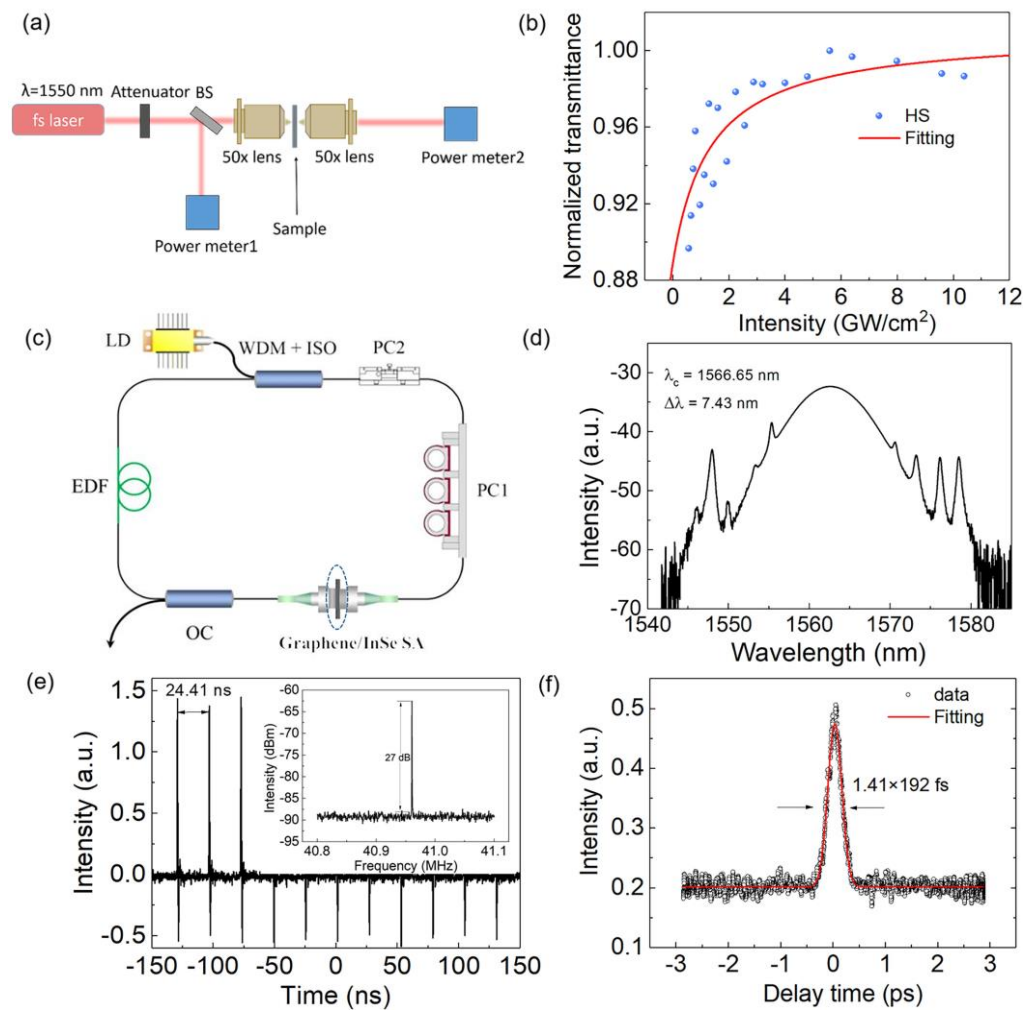

**Figure 4.** (a) Schematic diagram of the micro I-scan setup. (b) Nonlinear saturable absorption curve of G/InSe HS. (c) Schematics of the mode-locked fiber laser based on G/InSe HS SA.. (d) Output optical spectrum based on G/InSe HS SA (The spectral resolution is 0.02 nm). (e) Oscilloscope trace of mode-locked laser state. Inset figure is the obtained RF spectrum with 100 Hz resolution bandwidth (RBW). (f) AC trace of mode-locked pulses after amplification and compression.

## 5. Conclusions

High-quality G/InSe HS material was prepared by ME and dry-transfer method, and the carriers transport across the G/InSe HS interface was systematically investigated by PL, KPFM and TA measurements. The relatively lower saturation intensity (~1.33 GW/cm²) and larger modulation depth (~12%)are obtained by nonlinear absorption measurement. Furthermore, mode-locked laser pulses with 192 fs pulse duration operating at 1566.65 nm is successfully generated by integrating the G/InSe HS into Erbium-doped fiber laser cavity. The obtained pulse width is superior than the naked graphene SA in the same laser cavity. The internal mechanism is the reduced rate of recombination, and fast interlayer charge transfer driven by the large built-in electric potential of G/InSe HS. Our work not only provides a deeper understanding on the dipole orientation related interface interactions of G/InSe HS, but also demonstrates its application prospects in the field of nonlinear photonics.

**Supplementary Materials:** The following supporting information can be downloaded at: www.mdpi.com/xxx/s1, Figure S1: Mode-locked laser pulse based on graphene SA; Figure S2: The TA signal of graphene and G/InSe HS with pump and probe laser wavelengths of 1350 nm and 1566 nm respectively. Figure S3: AFM image of graphene and InSe.

**Author Contributions:** J.L. prepared the devices; J.L., Y.C, Y.L. and H.Z. performed the TA measurements and analyzed the experiment data. J.L. and L.W. performed the mode-locked laser experiments; J.L. and L.L. wrote the paper; L.L. and L.T. made revisions and finalized the document. All authors have read and agreed to the published version of the manuscript.

**Funding:** This research was supported by the National Key Research and Development Project of China (2018YFB2200404), the National Natural Science Foundation of China (91950205 and 61635009), the Fundamental Research Funds for the Central Universities (2019FZA5003) and the General program (11774308).

**Data Availability Statement:** The data can be requested from corresponding author upon reasonable request.

**Acknowledgments:** The authors would like to thank Yudong Cui for the suggestions of mode-locked laser experiment.

**Conflicts of Interest:** The authors declare no conflict of interest.